# Modeling structural change in spatial system dynamics: A Daisyworld example

Neuwirth, C.[a], Peck, A.[b], and Simonovic, S.P.[b]

[a]Doctoral College GIScience, University of Salzburg, Salzburg, Austria; corresponding author
Email: christian.neuwirth@stud.sbg.ac.at
Tel.: 0043 650 8424928
[b]Department of Civil and Environmental Engineering, University of Western Ontario, London, Canada

## Abstract

System dynamics (SD) is an effective approach for helping reveal the temporal behavior of complex systems. Although there have been recent developments in expanding SD to include systems' spatial dependencies, most applications have been restricted to the simulation of diffusion processes; this is especially true for models on structural change (e.g. LULC modeling). To address this shortcoming, a Python program is proposed to tightly couple SD software to a Geographic Information System (GIS). The approach provides the required capacities for handling bidirectional and synchronized interactions of operations between SD and GIS. In order to illustrate the concept and the techniques proposed for simulating structural changes, a fictitious environment called Daisyworld has been recreated in a spatial system dynamics (SSD) environment. The comparison of spatial and non-spatial simulations emphasizes the importance of considering spatio-temporal feedbacks. Finally, practical applications of structural change models in agriculture and disaster management are proposed.

## 1. Introduction

System dynamics (SD) is a computer simulation problem-solving approach with a foundation in concept of system feedbacks with the purpose of gaining insight into real-world system behavior (Forrester, 1969). This approach was first introduced by Jay Forrester in the mid-1950s as a framework for the conceptual representation as well as the quantitative modeling of economic systems (Radzicki & Taylor, 1997). *The paradigm of System Dynamics itself assumes that things are interconnected in complex patterns; that the world is made up of stocks, flows, and feedback loops; that information flows are intrinsically different from physical flows; that nonlinear processes and delays are important elements in systems; and that behavior arises out of system structure (Meadows, 1989, p16).*

Besides economical and business management problems, a relatively large number of SD models have already been applied to study environmental processes, especially in the field of hydrology and water resource management (e.g. Ahmad & Simonovic, 2000; Gastélum, Valdés, & Stewart, 2010; Khan, Yufeng, & Ahmad, 2009). The main purpose of these models is to provide insight into non-linear system behavior for the ultimate purpose of assisting informed decision making by stakeholders and policymakers.

However, SD was originally developed as an approach for modeling non-spatial systems. Efforts to integrate spatial modeling capacities into SD resulted in the development of sophisticated software products as for instance Spatial Modeling Environment (SME) pioneered by Maxwell and Costanza (1997) or SIMILE presented in Muetzelfeldt and Massheder (2003). Moreover, implementations such as SIMARC (cf. Mazzoleni et al., 2003) or 5D (cf. Mazzoleni et al., 2006) were proposed to address shortcomings of SD in modeling spatial processes. Concepts as well as the availability of mature software products enabled the application of SSD to a number of case studies in different fields such as ecosystem modeling, hydrology or invasive species control (e.g. Ahmad & Simonovic, 2004; BenDor & Metcalf, 2006; Voinov et al., 2004).

The main focus of these models is on the simulation of spatial diffusion processes. The general principle of this model type is that elements spread from areas with high concentration to areas with lower concentrations. In contrast, the simulation of evolving spatial structures has been widely neglected so far in SSD. The term structure may be defined as *"the arrangement of and relations between the parts or elements of something complex" (*Oxford Dictionaries, 2014). A structural change model deals with the temporal evolution of spatial structures and the processes driving this evolution. An example of this type of model is a land cover model. "*Land cover is defined by the attributes of the earth's land surface and immediate subsurface" (Lambin, Geist & Lepers, 2003, p213)*. Natural and anthropogenic processes modify the spatial arrangement of these attributes over time.

So far, this type of spatio-temporal system has been modeled by only very few modelers in SSD. Furthermore, models currently in existence either represent structures merely as relational space (e.g. BenDor, 2012) or neglect feedback between processes and structures (e.g. Lauf et al., 2012). Therefore, the objectives of this article are (i) the introduction of concept and techniques to address these shortcomings in structural change models (ii) and to verify the impact of process-structure feedback on a system's behavior.

This is evaluated based on the Daisyworld model, introduced by Watson and Lovelock (1983). Daisyworld is a fictional system, which serves as a simple parable to show effects of feedback between life and its environment (Wood, Ackland, Dyke, Williams, & Lenton, 2008). In this example, the close linkage of the biota to the atmospheric environment causes self-regulation. In order to determine the effects of structural feedbacks on the system's behavior, simulation results of two different Daisyworld versions with and without spatially explicit landscape structures are compared.

A more detailed explanation of the Daisyworld model and its basic idea is given in section 4. In the following sections established SD and SSD concepts are explained and a selection of current SSD models in literature is discussed. This is followed by a methodological discussion on the technical implementation and the nature of structural change models in SSD. Subsequently, the spatial and non-spatial Daisyworld simulations are explained in detail. The simulation results are presented and discussed. Finally, conclusions are presented and suggestions are made for future applications.

# 2. Spatial System Dynamics

2.1 Concept

System dynamics (SD) combines mathematics and computer simulation to explore the behavior of real-world systems, relationships and processes over time. The SD approach is effective for formalizing system structure, providing a better understanding of what drives system behavior and examining future dynamics based on a given set of assumptions. In this way, modeling system behavior over time can provide insight into significant relationships, reveal patterns, expose sources of undesirable system behavior, and help avoid unforeseen consequences of future policy implementations. Two important properties of the approach include system *model structure* and system *simulation*.

System structures can be conceptualized by means of a causal loop diagram (see Fig. 1, a). Arrows indicate relationships between system elements and can expose feedback loops inherent in the system. The relationship between variables can be characterized by assigning a polarity (positive or negative) based on the direction of the change. For example, a negative polarity indicates that the change of a variable in one direction causes the second variable to change in opposite direction. A positive polarity, however, means that variables change in the same direction. The behavior of a particular loop is driven by system structure and the polarities of variables in a particular loop. Negative feedback may be considered "goal seeking" or "stabilizing" and conversely, positive loops are often considered "diverging from equilibrium" or "destabilizing"; and thus negative feedback is generally more desirable than positive feedback.

The conceptual causal loop diagram serves as a blueprint for the implementation of an executable stock and flow model (see Fig. 1, b). This type of model is comprised of stocks (accumulations or depletions), flows (rates of change), delays (time lags) and feedbacks (reciprocal relationships). In order to simulate behavior, the SD model uses mathematics to describe and relate model stocks and variables to each other. Stocks are mathematically represented as integrals, which provide a form of memory to the system. Flows are rates of change expressed as a set of non-linear first-order differential equations.

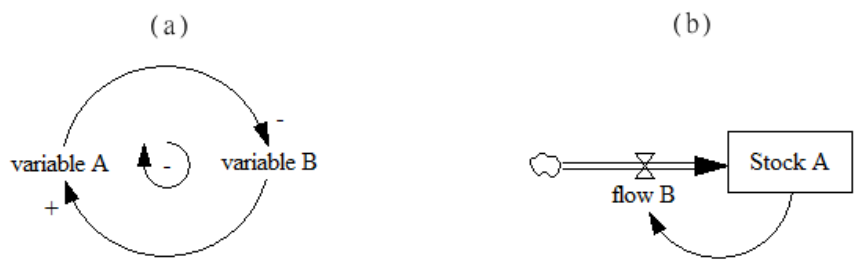

**Fig. 1.** Schematic representation of a (a) causal loop diagram; (b) stock and flow diagram.

The field of SSD modeling aims at an integration of space in SD models. There is a need for spatial and temporal dynamic simulations, as the non-linear behavior of highly complex systems cannot be intuitively forecasted. Real-world systems are highly dynamic in both time and space and unanticipated, even counter-intuitive behavior can often arise in complex systems (cf. Forrester 1971), because oftentimes *"cause and effect are distant in time and space" (Sanders & Sanders, 2004, p3)*. Therefore, in order to more accurately simulate behavior of real-world systems, it is necessary to consider both the short and long-term dynamics of system behavior including the impacts and interactions between time and space. This ultimately assists in assessing the effectiveness of policy decisions and helps identify those actions which would help achieve desired goals. Although SD research is trending in this direction, *"the spatial dimension has not received a great deal of attention in system dynamics modeling. An intensive literature review showed that there are only a number of articles dealing with this subject" (Sanders & Sanders, 2004, p9)*.

## 2.2 Architecture of SSD Models

The question of how to represent physical space in a model and how to link this representation to a dynamic SD model is crucial in the field of SSD modeling. Raster or vector data models with Graphical Information Systems (GIS) are commonly used to represent spatial environments. The linkage of SD variables to a spatial GIS data model includes both data association as well as semantic association (Zhang, 2008).

Data association refers to technical coupling and the way of managing the mutual exchange of data between SD and GIS. De Smith et al. (2007) distinguish between loose, moderate and tight coupling approaches for the connection of two standalone computer applications by data transfer. In a loosely coupled SSD model, functions operate asynchronously within each system. For instance, the GIS may be used to preprocess spatial inputs, which are then passed to a SD model; thereafter, execution results are returned to the GIS for visualization and additional analysis. A moderately coupled SSD model enables an indirect communication of systems by means of shared access to a database. Finally, tight coupled applications are characterized by the synchronous operation of systems, allowing direct inter-system communication during program execution. This is achieved by invoking commands from both SD and GIS in a single script. Longley et al. (2005) suggest the term "embedded" to classify solutions which fully integrate modeling functionality in one independent piece of software.

Since variables exhibit different manifestations within SD and GIS (Zhang, 2008), a semantic association also needs to be defined for the SSD model. For instance, raster cells or vector features (point, line or area objects) of the GIS data model may be linked to SD stocks. In this way, dynamically changing values of the stock element in SD can be transferred to the spatially decomposed simulation environment in GIS and vice versa. Due to structural similarities, two-dimensional stock arrays are often associated to GIS rasters (e.g. BenDor & Metcalf, 2006, Ahmad & Simonovic, 2004). These multidimensional arrays are provided as a data type with most SD software applications. One significant advantage is the spatial topology (relative position of array elements) inherent in this data type. SSD applications based on vector data models are usually lacking information on the relative position of object features.

In addition to structural peculiarities, the type of spatial representations used in SSD models strongly depends on the processes being modeled. For instance, the continuous raster view tends to work best in describing physical quantities (Goodchild, 2005). Huggett (1993), for instance, introduced a SSD model of nitrogen accumulation based on terrain data represented by means of a spreadsheet raster. In this way, the spreadsheet provides the SD model with terrain information being used for the derivation of nitrogen flow directions. The relative positions of raster elements (neighborhood conditions), which is required to determine nitrogen flow directions, is defined in the stock array. The link to the spreadsheet is provided as a built-in function of the SD software and allows for spatial results to be displayed in two-dimensions.

A similar model was implemented by BenDor and Metcalf (2006) for the simulation of invasive species spread using the software SME. SME enables the incorporation of a generic SD model into a spatial array. The semantic association is equivalent to the nitrogen accumulation model. Stock variables are replicated and associated to two-dimensional raster representation of space. However, in contrast to the model introduced by Huggett, more complex neighborhood conditions are considered. The propagation of bark beetles is modeled as a function of beetle density in an origin raster cell and the distance to a destination raster cell. Moreover, the migration of beetles to adjacent cells is restricted by a maximum migration distance which is defined prior to the simulation. Conventional GIS is only used in this model to initially parameterize spatial arrays and to visualize spatial data in a raster format.

In contrast to SSD models with spatial extensions, SSD models based on GIS coupling benefit from a comprehensive set of spatial modeling and data management functionality. Ahmad and Simonovic (2004), for instance, coupled SD to GIS for the simulation of damages caused by floods in the Red River basin (Canada). Similar to the models presented so far, diffusion processes were modeled based on raster data. SD captures spatio-temporal dynamics of discharge depths, which are sent to a corresponding GIS raster. Damages to the infrastructure affected by the modeled flood is reported and visualized in GIS. The data association was achieved by passing data from SD to GIS through a spreadsheet, using the dynamic data exchange (DDE) protocol. However, the authors concluded that this coupling approach doesn't enable a fully automated update of results in GIS.

Zhang (2008) overcame this problem by the integration of SD and GIS functionality in a Universal Development Environment (UDE) which serves as middleware to tightly couple these systems. In this way, the coupling can be achieved without producing intermediate data. This model was applied to the spatio-temporal simulation of contaminant propagation in the Shoghua River, China. The process is modeled on a raster-basis in SD and visualized in a customized component GIS. In contrast to Ahmad and Simonovic's flood model, a one-dimensional approach was selected for simulation of the pollutant concentration along the river course.

Compared to raster-based simulations introduced so far, vector approaches are underrepresented in the field of SSD modeling. Lowry and Taylor (2009), for instance, proposed a technical solution for coupling SD to Google Earth, which enables the update of KML (vector format) color attributes based on SD variables. Similar to the model on pollution propagation by Zhang, the systems were tightly coupled by invoking commands in a separate script. The definition of relative position of vector polygons was neglected, since no interaction between vector polygons was foreseen.

The preference of modelers for raster-based approaches and a one to one raster cell-to-stock ratio is probably associated with the predominant use of SSD applications for modeling continuous diffusion processes. One major limitation of this concept is the limited size of dimensional arrays in established SD software. This is because the array data type was originally designed for non-spatial models. This limits the applicability of SSD, especially in the fields of hydrology and meteorology where extensive arrays and high resolutions are required. In addition to this software limitation, SSD models usually neglect mutual interactions between SD and GIS. Typically, communication is only in a single direction from SD to GIS, which commonly serves purely as a visualization tool.

# 3. Methodology

An application of SSD for modeling evolving spatial structures requires concepts and techniques which partially differ from those used in diffusion models. The following sections constitute a first attempt to define the structural change model type and to describe a general framework for an implementation in SSD.

## 3.1 The structural change model type

Structural change models are characterized by the following features: 1. A change of spatial structures concerns a *qualitative change* of spatial objects as, for example, a change from one land cover type to another. 2. Processes interact with space and lead to observable patterns and structures in space (cf. Getis & Boots, 2008). 3. Structures determine how processes come into effect as, for instance, studied in landscape ecology (Turner, 1989).

This implies that there exists a feedback loop between processes and structures which has to be considered in structural change models. In SD, feedback elements are non-spatial and expressed numerically. In the case of the structural change model however, landscape topologies are involved as a spatial feedback element. This type of feedback may be termed *structural feedback*.

The characteristics introduced can be illustrated by means of a simple grassland farming example. In this example, cultivation activities such as harvesting, transport and storage of forage constitute processes. Spatial structures are given by a transportation network of farm lanes. The intensity of the ongoing farming process can be quantified by the forage gained per area. This intensity affects the infrastructure, since maintenance of farm lanes depends on their usefulness. The quality of infrastructure, however, also affects the intensity of farming as it can be assumed that well exploited areas are farmed more intensively and vice versa. This system can be formalized in a simple non-spatial SD model (see Fig. 2).

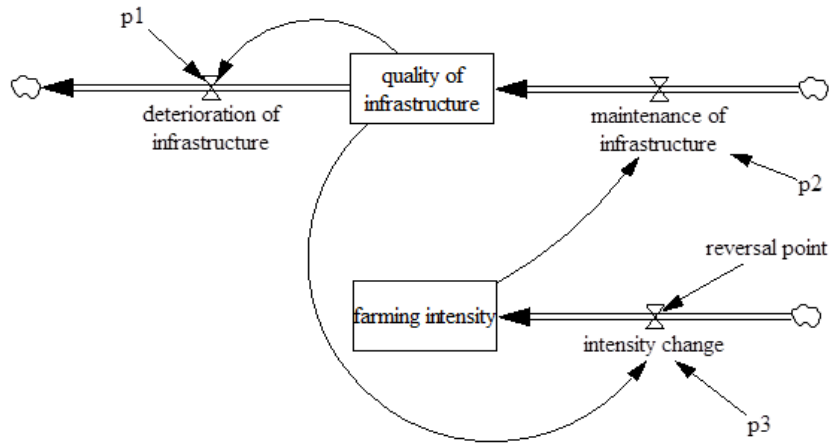

**Fig. 2.** SD model on farming intensities and infrastructure

Causal dependencies are represented by means of proportionality factors (factors p1, p2 and p3). In this example, the quality of infrastructure deteriorates proportional to itself, whereas rate of maintenance is proportional to the level of farming intensity. Farming intensity is dependent on the level of infrastructure quality as a function of the rate of intensity change.

In the presented realization of this system, the infrastructure is rated based on a single quantitative measure. In a spatial model, however, infrastructure may actually be a collection of various different infrastructure types (e.g. roads, bridges, buildings, other). In addition, the infrastructure may be further described by defining characteristics (e.g. if the infrastructure is roads, they may be further categorized as paved or unpaved). Farm road types are modified as a function of farming intensities, which may be considered a result of spatially distributed (multiple farmers) or lumped farming activities (farming as a global process). This enables a derivation of travel times from an infrastructure map, which can be returned as variable to update farming processes. This two-way interaction provides improved realism of structural system feedbacks and simulation results. The way structures and processes are integrated in this model type is explained in section 3.2.

## 3.2 Semantic association

Semantic association in SSD concerns the way non-spatial SD variables are linked to the spatial environment being modeled. In the presented approach, stock variables are related to the size of structural elements. This may be the area of land cover patches (see Fig. 3) or the length of different types of transportation infrastructure.

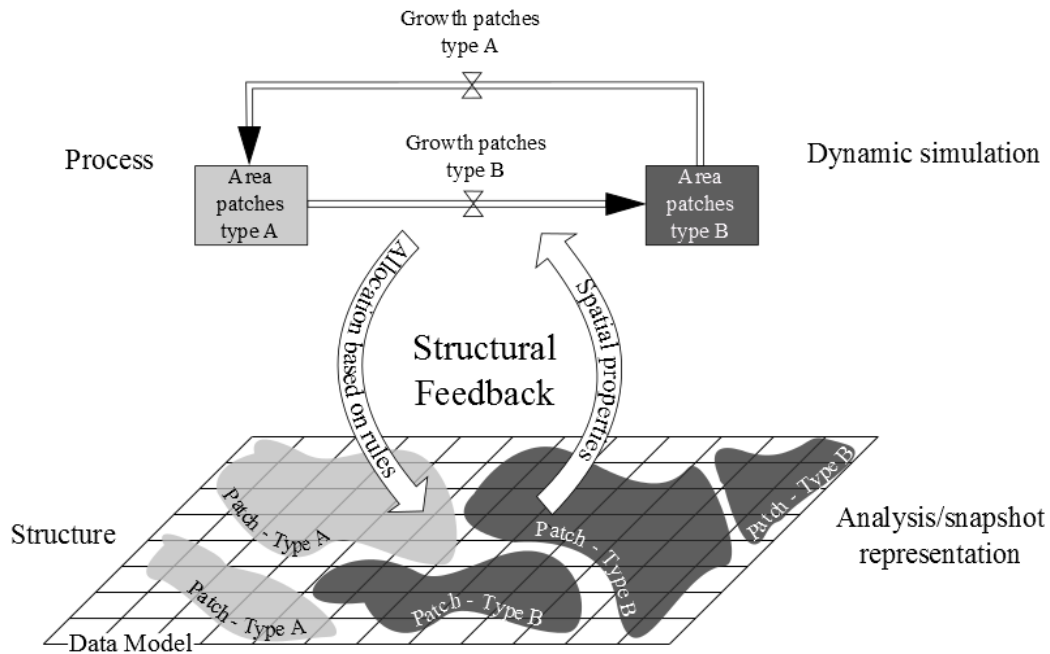
**Fig. 3.** Association of process (time) and structure (space) in a structural change model

These structural elements are related to the stock values and updated based on spatial allocation rules. For many applications, the growth of an object is validly described by assuming least cost or gravitational rules. For instance, a reduction in object size may be based on the principle of cost optimization. The example mentioned in section 3.1 related to agricultural infrastructures similarly aims at coverage which most efficiently services the most profitable areas.

The raster data model lends itself nicely to this allocation task. Raster cells are assigned to stocks and ownership is exchanged between them throughout the simulation. Accordingly, stocks are linked to multiple raster cells. Thus, the proposed method breaks the common one to one raster cell-to-stock ratio of established SSD applications (see also section 2.2).

Conversely, the relationship of stocks to spatial objects is application dependent. A stock may be related to a single region with a closed boundary. However, a stock could also be related to multiple, non-adjacent zones like land use patches of the same type (see Fig. 3). The relationship of regions or zones of different type and their interactions are defined in the stock and flow diagram.

Once the stock variable has a spatial counterpart, spatial implications such as the resistance to further growth can be derived from georeferenced data layers. Dynamically changing spatial properties are analyzed and transferred back as variables to the SD model. This link between evolving objects at different points in time acts as a spatio-temporal (structural) feedback loop in the system. The technical implementation of the bidirectional interaction between process and structure is described in the following section.

3.2 Technical implementation

In order to lay the foundation for managing spatial complexity in SD, the tight coupling of traditional SD software with GIS is proposed. The following section discusses the approach taken to develop a custom coupling program and the conversion of time-continuous quantities to spatial representations and vice versa. VENSIM software provides the SD modeling and simulation capabilities and ArcGIS software introduces the spatial element to the model. Despite the fact that both VENSIM and ArcGIS are proprietary software packages, they are also mainstream programs used widely in the SD and GIS communities. The coupling application was developed using ArcGIS version 10.1; VENSIM DSS version 5.10; and Python version 2.7.2. This software is required for developing, modifying and running the coupling program.

The custom coupling application uses DLLs (*.dll*), Python libraries (*.lib*), packaged VENSIM models (*.vpm*), and raster-based GIS databases (*.mdb*) to create an environment for dynamically linking VENSIM model simulation and GIS. VENSIM Dynamic Library Links (DLLs) are used to access VENSIM models and execute commands via the VENSIM DLL to give control over VENSIM model settings, variables and simulation parameters.

The coupling application operates differently than a regular VENSIM simulation; it does not just execute a pre-determined simulation sequence; the coupling application actually uses VENSIM's Game command to provide control over model simulation. First, the coupling program initializes libraries, DLLs, save drives, and other necessary start functions; loads a published model file (*.vpm*) and loads a GIS raster file (*.mdb*). The user is then asked to define simulation settings, following which, pertinent spatial data is extracted from raster files and sent to variables in the VENSIM simulation model. Simulation then begins by executing a single time step in the SD simulation model. Upon completion of a time step, the simulation data is used as input into the spatial analysis component of the program. Spatial analysis is completed for a single time step and results of spatial analysis are sent back to the SD model. The same simulation process continues until the specified final time is reached and the simulation ends. During a simulation, the coupling program tracks simulation progress and calls the VENSIM DLL to sketch temporal results as graphs in real-time and uses ArcGIS to generate a semi-continuous sequence of maps.

Tight coupling is achieved by executing scripts in a stand-alone Python program. The Python language is one that is freely available and quite popular in the programming community. The logic behind selecting Python as the sole programming language is for code reusability, consistency and program transferability. Despite the fact that VENSIM software has java, C++ and Visual Basic support tools, Python was selected to develop the custom coupling program for reasons of software compatibility, cross platform functionality, open source development and intended future use.

Python has an extensive collection of useful packages which facilitate program development. The Python packages imported into the custom coupling program include: *Numpy* (for scientific computations and array operations); *ArcPy* (for getting access to ArcGIS functionality); *matplotlib* (for visualizing simulation output); and ctypes (a foreign function library for compatibility with C data types). The custom coupling application has been developed using the *Numpy* package to replace the equivalent spatial analysis functions of *ArcPy*. This offers more flexibility for a future adaptation of raster operations. Thus, *ArcPy* provides spatial references as well as the interface to the ArcGIS database whereas the spatial analysis is conducted in *Numpy*. In the further course of the paper the spatial component of the model is summarized by the term "GIS".

One of the unique elements of the approach and specific application in this paper is the synchronous data flow between SD and GIS. To effectively have SD simulate system processes and GIS to perform spatial analysis requires bidirectional synchronous operation between SD and GIS software. An example of this synchronous operation loop is presented in Fig. 4.

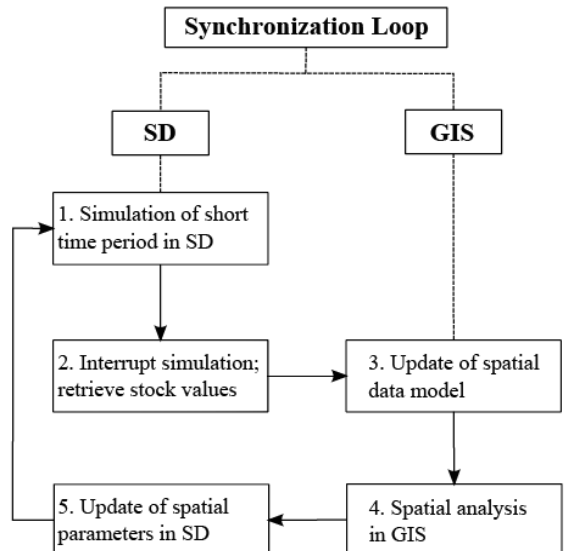

**Fig. 4.** Schematic representation of synchronized operations between SD and GIS

In the SD simulation model, temporal memory is preserved between time steps while spatial analysis is performed. In effect, generating snapshots of changes in spatial structure over time. This differs from the simulation of diffusion models and diffusion processes, where no interruption of the SD model simulation is required. The main reason for this is because simulation of diffusion models does not require data to be returned from GIS to the SD simulation. Thus, diffusion systems may operate asynchronously in a SSD model.

This coupling approach could in theory be used for any number of potential SSD applications which involve spatio-temporal processes. In the following section, Daisyworld is described and expanded to serve as an example of implementing the tight coupling approach.

# 4. Daisyworld

Daisyworld is a fictitious environment introduced by Watson and Lovelock (1983) to describe biological homeostasis of the global environment. The term homeostasis originated in the field of physiology and refers to *"the ability of living beings to maintain their own stability" (Cannon, 1929, p1)*. The intention of the original Daisyworld model was to illustrate that homeostasis could also be observed in environmental systems as described by the Gaia hypothesis (Lovelock and Magnus 1974).

In the Daisyworld model a fictitious planet accommodates three land cover types classified as black daisies, white daisies and fertile soil (known as "bare ground" in the original Daisyworld model). A distant sun emits luminosity and warms up the planet's surface. Due to the difference in albedo between black and white daisies, local temperatures next to the daisy patches differ from average planetary temperatures. Black daisies cause a local heating, whereas temperature next to areas vegetated by white daisies falls below average planetary temperatures. Since both types of daisies reach a growth optimum of 100% at a temperature of 22.5°C, localized heating leads to beneficial conditions for black daisies in a Daisyworld with average planetary temperature below the optimum temperature. Conversely, local cooling favors white daisies in an environment where temperature is above this threshold.

This mechanism counteracts changes of the sun's luminosity by means of two negative feedback loops. Increasing luminosity leads to higher temperature which favors the growth of white daisies; which, in turn, decreases the temperature in Daisyworld due to a lower planetary albedo. The same principle applies to decreasing luminosity which causes a spread of black daisies. In this way, type of land cover acts as a stabilizing component within the system and can be considered a driving factor of homeostatic control in Daisyworld.

**Table 1.** Daisyworld equations (Watson & Lovelock, 1983)

| # | Description | Equations |
|---|---|---|
| (1) | Planetary Albedo | $A = \alpha_g A_g + \alpha_b A_b + \alpha_w A_w$ |
| (2) | Planetary Temperature | $T_e = \sqrt[4]{(SL(1-A)/\sigma)} - 273$ |
| (3) | Local Temperature | $T_{b,w} = q'(A - A_{b,w}) + T_e$ |
| (4) | Growth Rate | $\beta_{b,w} = 1 - 0.003265(22.5 - T_{b,w})^2$ |
| (5) | Area Change | $d\alpha_{b,w}/dt = \alpha_{b,w}(x\beta_{b,w} - \gamma)$ |

Planetary albedo A; area α; subscripts: fertile soil g, black daisies b, white daisies w; planetary temperature $T_e$ in [°C]; solar input S, 917Wm$^{-2}$; dimensionless measure of the luminosity of Daisyworld's sun, 0-1; Boltzmann constant, 5.67032E$^{-8}$; proportional constant q', 20°C, daisy growth rate β; area of fertile soil x; death rate per unit of time y, 0.3

Processes involved in this system are described by the equations in Table 1. The albedo of the planet is calculated as an area-weighted average of the albedos from the three surfaces (see Table 1, equation 1). Planetary temperatures are based on the Stefan Boltzmann law, which determines temperatures as a fraction of the absorbed energy input. For the sake of simplicity, the planet is treated as a planar surface with uniform energy input (see Table 1, equation 2). Local temperatures in the vicinity of black and white daisies are expressed by the difference in daisy and planetary albedos (see Table 1, equation 3). Daisies grow as a parabolic function respective to their local temperature (see Table 1, equation 4). Moreover, growth is affected by the globally available fertile soil patches not covered by daisies. Growth rates linearly decrease as available fertile soil patches decrease. In addition, it is assumed that a fraction of daisies die off each time step (see Table 1, equation 5).

4.1 Spatial Daisyworld

Ford (2010) implemented Daisyworld as a non-spatial, dynamic model in SD. The proposed SSD Daisyworld model fulfills the aforementioned criteria of process-structure interaction defined for structural change models (see section 3.1) and may therefore be considered an appropriate example in demonstrating spatio-temporal effects for this model type.

In order to consider a spatial Daisyworld model, there were two significant modifications made to the original model: (i) spatialization of Daisyworld; and (ii) the introduction of *barren land* as a fourth land cover type. Otherwise, all other settings remained consistent to the original model. Barren land simulates the impact of spatial barriers which prevent daisies from further propagation. The major intention of introducing this supplementary land cover type is to put more emphasize on spatial implications such as differences in conditions regarding location. The additional land type is intended to reinforce effects of spatial landscape structures on the simulation. Barren land is considered as immutable and unproductive land in the Daisyworld simulation environment, so to accommodate this, a supplementary constant was introduced into the Daisyworld SD model. An additional assumption made in the spatial Daisyworld model is that barren land has an albedo value of 0.5. The initial values and settings used in the spatial Daisyworld model are presented in Table 2, according to the numeration seen in Fig. 5.

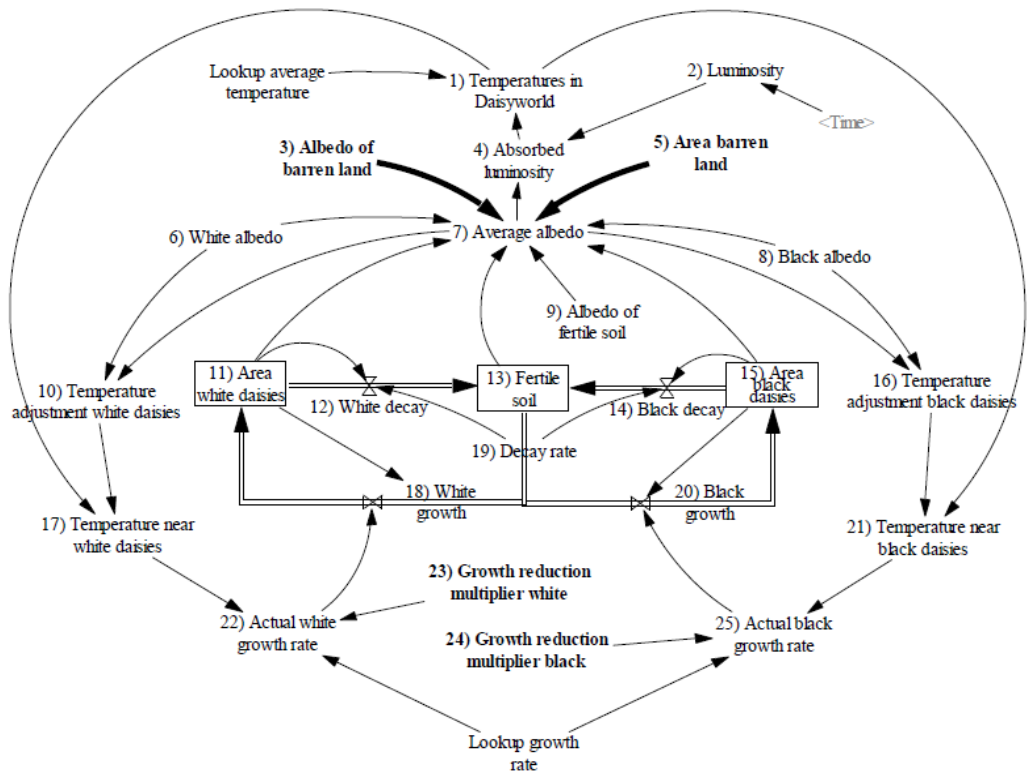

**Fig. 5.** Daisyworld as stock and flow diagram; variables which have been added to the original model are shown in bold

The area of different land cover types such as the area of black daisies (see Fig. 5; No. 15), or white daisies (see Fig. 5; No. 11) are represented by stocks in the SD model. In order to initialize the model, stock values are related to a randomly generated Daisyworld landscape raster. This is implemented in such a way that the areas of respective land cover types in the spatial raster correspond to the initial stock values.

The growth of daisy populations in SD requires updating in the spatial representation based on a set of allocation rules. It is assumed that daisies only grow next to daisies of the same species. Therefore, in a first step fertile soil cells sharing a border to a daisy cell of the respective species are selected. From this pool of candidates cells are randomly selected to allocate daisies. This also implies that vegetated areas grow laterally, provided that sufficient fertile soil is available. Alternatively, this simple allocation rule may be extended in the future by enabling growth within a distance buffer to daisies of the same species.

**Table 2.** Detailed description of the Daisyworld stock and flow elements (based on descriptions in Ford, 2010); modifications to this model are referenced by footnotes

| #      | Type      | Init. or const. value | Condition or equation                      |
|--------|-----------|-----------------------|--------------------------------------------|
| (1)    | Function  | -                     | Equation 2                                 |
| (2)    | Condition |                       | Time dependent cond.: IF THEN ELSE(Time<x,y,z) |
| (3)[a] | Constant  | 0.5                   | -                                          |
| (4)    | Variable  | -                     | Luminosity*(1-Average albedo)              |
| (5)[a] | Constant  | Scenario dependent    |                                            |
| (6)    | Constant  | 0.75                  | -                                          |
| (7)[b] | Variable  | -                     | Equation 1                                 |
| (8)    | Constant  | 0.25                  | -                                          |
| (9)    | Constant  | 0.5                   | -                                          |
| (10)   | Variable  | -                     | (20)*(Average albedo-White albedo)         |
| (11)[c] | Stock | Scenario dependent | White growth-White decay |
| (12) | Flow | - | Area white daisies*Decay rate |
| (13) | Stock | Scenario dependent | Black decay+White decay-Black growth-White growth |
| (14) | Flow | - | Area black daisies*Decay rate |
| (15)[c] | Stock | Scenario dependent | Black growth-Black decay |
| (16) | Variable | - | (20)*(Average albedo-Black albedo) |
| (17) | Variable | - | Temperatures in Daisyworld+Temperature adjustment white daisies |
| (18) | Flow | - | Area white daisies*Actual white growth rate |
| (19) | Constant | 0.3 | - |
| (20) | Flow | - | Area black daisies*Actual black growth rate |
| (21) | Variable | - | Temperatures in Daisyworld+Temperature adjustment black daisies |
| (22) | Function | - | Equation 4 |
| (23) | Spatial variable | - | Local availability of fertile soil for white daisies |
| (24) | Spatial variable | - | Local availability of fertile soil for black daisies |
| (25) | Function | - | Equation 4 |

[a] New land type barren land
[b] See equation 6; barren land is involved as a fourth land type in the calculation of planetary albedo
[c] See table 1, equation 5; the globally defined growth factor $x$ which depends on the area of fertile ground in the original version is replaced by a spatially explicit growth reduction multiplier $D_{b,w}$ (see also modified equations 7 and 8)

In addition to the temperature dependent growth of daisies, there is a reduction in daisy area as they die. It is assumed that 30% of daisies die in each time step. This acts as a negative feedback loop in the SD model which triggers an increase in decay as the daisy area increases (see Fig. 5, No. 12 and 14). In order to spatially allocate daisy decay, a timestamp is provided to every raster cell that indicates the age of daisy growth. The age of the daisies is used as criteria in determining which raster cell subsequently experiences daisy death. The raster cell with the oldest daisies are sequentially selected and updated to reduce the amount of daisy growth. When a raster cell's daisies die, the cell returns to fertile soil type. This continues until spatial representation corresponds to the respective stock values.

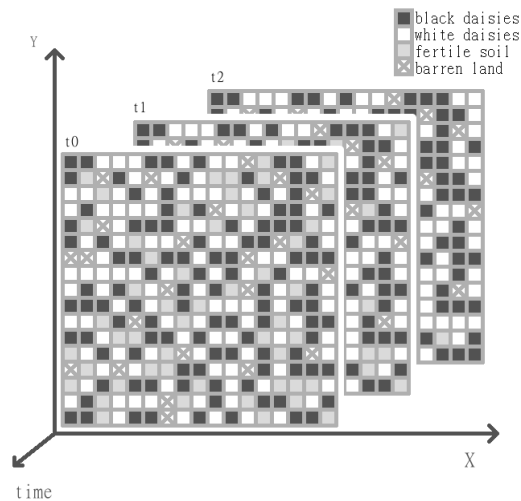

**Fig. 6.** Schematic representation of spatio-temporal landscape changes in Daisyworld

Due to the dynamic evolution of the landscape (see Fig. 6), the opportunities for a species to spread changes as a function of soil availability. The growth reduction control in the original version assumes that the remaining soil is equally shared between the two species. The scarcity of soil is imposed as a globally defined factor by reducing the growth rate of both species as a linear function of fertile soil (see Table 1, equation 5). For instance, if 50% of Daisyworld is fertile soil, growth rates of both species are multiplied by 0.5. In a spatial Daisyworld however, the distribution of soil resources is a matter of locally available fertile soil patches.

Thus, the original model was supplemented with an individual growth reduction multiplier for each type of daisy (see Fig. 5; No. 23 and 24). The growth reduction multiplier reflects the spatial proximity of each species to fertile soils. The edge length between landscape elements is important for the growth of species (Turner, 1989). This is considered by counting the number of shared borders between daisy cells and fertile soil cells. In theory, each daisy cell can be adjacent to eight fertile soil cells. The proportion of actual adjacencies to the maximum number of adjacencies is used as a proxy of available expansion areas for each daisy species.

In order to sequentially search the raster for adjacencies, a 3x3 moving window is used. This procedure is repeated each time the Daisyworld landscape changes to update the SD model with new information. In order to take these modifications into consideration, equations 1 and 5 (see Table 1) were adapted as follows:

$$A = \alpha_g A_g + \alpha_b A_b + \alpha_w A_w + \alpha_{barr} A_{barr} \qquad (6)$$

$$d\alpha_{b,w}/dt = \alpha_{b,w}(D_{b,w}\beta_{b,w} - \gamma) \qquad (7)$$

The planetary albedo was calculated including the area and albedo of barren land. Furthermore, the area of fertile soil in equation 5 was replaced by the growth reduction multiplier for each species (b, w) calculated by

$$D_{b,w} = G_{b,w}/C_{b,w}8 \qquad (8)$$

where $G_{b,w}$ is the number of fertile soil raster cells next to black or white daisies and $C_{b,w}$ is the number of raster cells occupied by black or white daisies.

# 5. Discussion of results

A spatial and a non-spatial version of Daisyworld are simulated and compared for various scenarios to verify the impact of structural feedback on system behavior. Additionally, the performance and functionality of the proposed coupling approach is illustrated. In order to make the two versions comparable, the non-spatial model is also supplemented with *barren land* as an additional land type.

For all scenarios, a change in luminosity was assumed which is balanced out by means of land cover adaptation in the system. Since the initial composition of the landscape effects the simulation results, multiple random landscapes were generated for the spatial simulations. Results generated for these landscapes were compared amongst each other and to the non-spatial simulation results. Furthermore, a stress test scenario was conducted by assuming disproportional black and white daisy areas. This scenario was intended to compare the resilience of spatial and non-spatial models under unfavorable conditions. In addition, luminosity was gradually increased over a longer time period in order to observe the continuous adaptation of both systems to a changing environment.

## 5.1 Comparison of spatial and non-spatial model

Both spatial and non-spatial Daisyworld simulations were run for 100 time steps (see Fig. 7). The luminosity value drops from 1 to 0.9 after fifty time steps. The land cover is composed of 100ha of each black and white daisies, 500ha of fertile soil and 300ha of barren land. Due to the initial random distribution of land cover types in the spatial model, multiple simulations were completed to identify effects of land cover distributions on the simulation results.

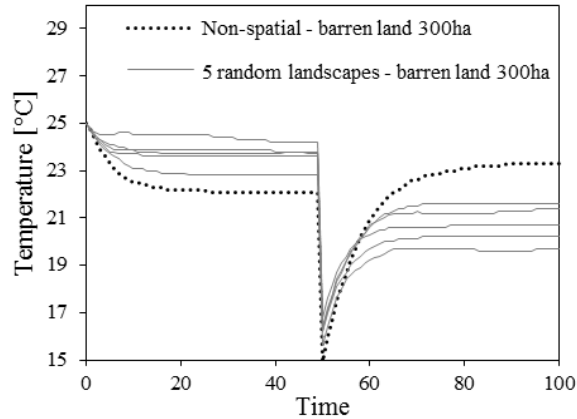

**Fig. 7.** Non-spatial simulation compared to 5 spatial simulations with randomly generated landscapes; model settings: black daisies 100ha, white daisies 100ha, fertile soil 500ha, barren land 300ha, luminosity drop from 1 to 0.9 after 50 time steps

Stable conditions are achieved after approximately 15 time steps since the onset of initial cooling. This cooling phase is much more pronounced in the non-spatial simulation than in the spatial simulation. The subsequent stable state is disturbed by the 10% reduction of luminosity. Whereas the non-spatial system is able to raise temperatures in this low radiation environment, the spatial system fails to completely resolve the outside disturbance to the system. This applies to all spatial simulations, regardless of their initial landscape distribution pattern. The variation in the results for five randomly generated landscapes, however, reveals considerable dependencies on land type distribution.

Nevertheless, the range of spatial results clearly differs from the non-spatial simulation. The main reason for these differences is a result of the quick response of daisy growth to changing luminosity and temperature in the non-spatial simulation which causes more pronounced temperature compensations.

This can be clarified by executing a stress test scenario. The scenario assumes a land composition of 190ha of black daisies, 10ha of white daisies, 500ha of fertile soil and 300ha of barren land. Luminosity drops from 1 to 0.8 after ten time steps. It can be assumed that these initial conditions cause an increase in black daisies and a decay of white daisies as a result of decreasing luminosity. Since white daisies are initialized with such a small area, the response of daisy growth to changing environmental conditions endangers the remaining areas.

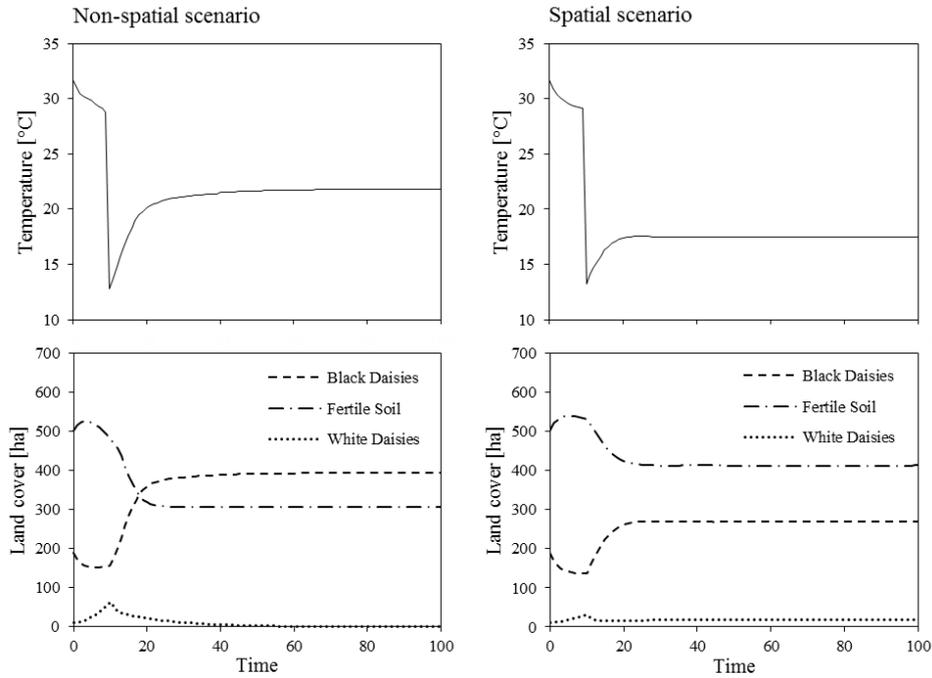

**Fig. 8.** Stress test scenario for a non-spatial and a spatial simulation; model settings: black daisies 190ha, white daisies 10ha, fertile soil 500ha, barren land 300ha, luminosity drop from 1 to 0.9 after 10 time steps

Initially, growth rates of white daisies are significantly higher in the non-spatial simulation than in the spatial simulation (see Fig. 8). However, the negative feedback between areas of white daisies and daisy decay causes higher decay rates as a result of decreasing luminosity after ten time steps in the non-spatial simulation. This leads to an extinction of white daisies and a loss of control mechanisms. Due to the slower decay of white daisies, this land cover can be sustained on a low level in the spatial Daisyworld. Therefore, although the non-spatial system is able to raise temperatures to a higher level, the spatial system is more resilient in the stress test scenario.

The main reason for this behavior may be attributed to the different daisy growth rates, which, in turn, are a result of the way available space for daisy propagation is defined in each of the respective models. In the non-spatial Daisyworld simulation, propagation of daisies doesn't require spatial proximity to fertile soil, and accessibility of fertile soil is explicitly considered in the spatial model by means of a growth reduction multiplier. This modification hinders daisy growth and crucially affects the response of the system to changing luminosity.

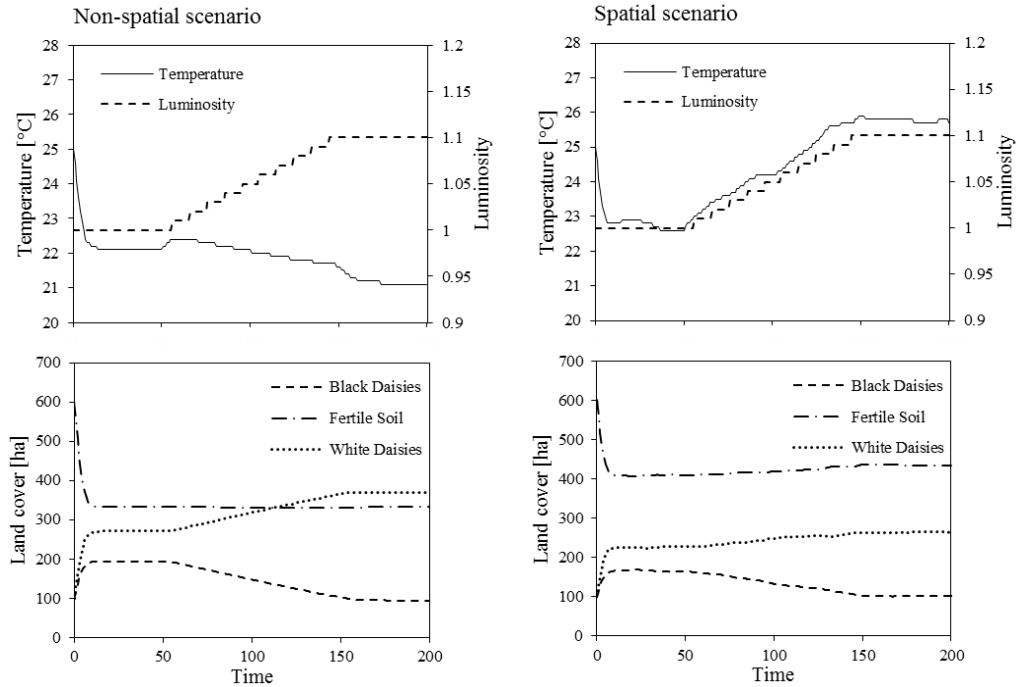

**Fig. 9.** Land cover response scenario for a non-spatial and a spatial simulation; model settings: black daisies 100ha, white daisies 100ha, fertile soil 500ha, barren land 300ha, luminosity is increased from 1 to 1.1 between time step 50 and 100

For instance, gradually increasing luminosity results in a period of cooling in the non-spatial system, since growth rates of white daisies unravel and even reverse effects of luminosity on temperatures (see Fig. 9). The spatial simulation however, reveals less of an impact of land cover adaptation on temperature reduction. Thus, the gradually increasing luminosity causes increasing temperatures. This is due to the restriction of suitable soils available for white daisies in the spatial simulation. The predefined growth restriction, allowing growth only next to daisies of the same species, leads to a local clustering of daisies. fertile soil which may be available elsewhere in Daisyland cannot be exploited if spatial barriers prevent daisy clusters from further expansion. This inhibits a growth magnitude required for sufficient cooling of Daisyworld. Land cover adaptation has a diminishing effect on temperatures, but there is no effective control on temperature when Daisyworld is simulated using a more realistic, spatial algorithm.

## 5.2 Sensitivity analysis and performance

The raster resolution in the spatial model was varied in order to illustrate effects on the model output. The settings used are equivalent to in the first spatial scenario. This simulation revealed significant and clear influence of the raster resolution on simulation results. An increase in spatial resolution can be associated with decreasing adaptive capacities of the system (see Fig. 10). Thus, results do not only depend on the composition of the landscape structure as shown in Fig. 7, but also on the way the landscape is represented in the data model.

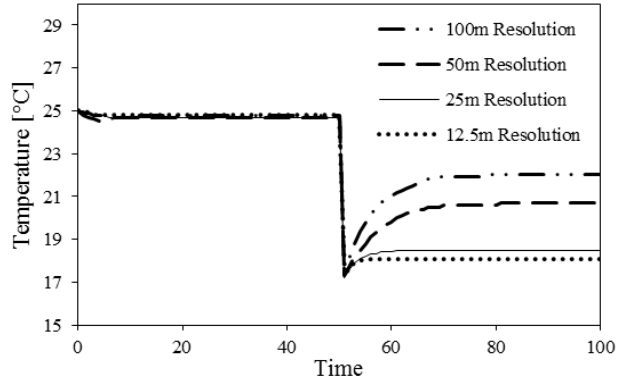

**Fig. 10.** Simulation of the same landscape with 4 different raster resolutions; model settings: black daisies 100ha, white daisies 100ha, fertile soil 500ha, barren land 300ha, luminosity drop from 1 to 0.9 after 50 time steps

The models' sensitivity to the spatial resolution is a consequence of results from the raster data model and model calculus. Consider a cluster of daisies to be composed of multiple daisy-type raster cells. There are daisy cells in the center of the daisy cluster, which are surrounded by other daisy cells. There are also daisy cells at the edge of the cluster that have a fraction of cell neighbors which are daisies and a fraction of cell neighbors which are fertile soil. A change in spatial resolution will modify the proportion of daisy raster cells on the edge of the cluster and the number of daisy cells at the center of the cluster, which biases the modeling results. As it turns out, if spatial resolution is increased there tends to be a more pronounced daisy growth reduction.

The sensitivity analysis was also used to test the computational performance of the coupling program. An increase in the spatial resolution partially leads to an overproportional rise of the processing time. The bisection of the resolution from 100 to 50 meters almost doubles the processing time from less than 10 seconds to about 17 seconds. If higher resolutions are used, the increase in processing times exhibits more of an exponential character. A simulation based on a 25m grid takes about 1 minute; more than 12 minutes are required to run the simulation with a resolution of 12.5m.

The increase in processing times results from spatial analyses which are performed on a raster basis. Increasing the time horizon (i.e. the number of time steps) of a simulation, reflects an increase in the number of interactions required between SD and GIS. However, this does not appear to produce any significant effects on processing times. Consequently, shortcomings in performance are not caused by the interaction between SD and GIS, but instead depend on Numpy raster analysis functionality invoked on the GIS-side of the program.

Furthermore, problems associated with the relatively limited number of stock variables in most SD products, as stated in section 2.2, are circumvented by the proposed approach. Due to the aggregation of raster cells, a smaller number of links is required between spatial entities and SD. This makes the structural change model type favorable for larger study areas.

Even though the number of links is smaller for this model type, interactions are bidirectional. Information exchange and update is not only for the purpose of visualizing spatial results, but introduces spatial (structural) feedback to the model. Therefore, modeling structural change in SSD makes higher claims against the quality and performance of links between SD and GIS. Thus, a link based on data in a script is preferable over an interaction via files. The usage of georeferenced raster data types provided with ArcPy in combination with multi-dimensional Numpy data types proved suitable for this task.

Next to coupling requirements, the need for a synchronous operation of functions (see Fig. 4) constitutes a further peculiarity of this application. The ability of the Vensim DLL to pause and continue the simulation at predefined points in time enabled an automation of this otherwise quite cumbersome task. The comparison of the performance of simulations with and without Vensim interruptions for 100 time steps revealed negligible differences in processing times (less than one second). This implies that the context switch doesn't produce a large overhead and indicates an overall reasonable performance of the proposed synchronization of operations.

# 6. Conclusions

The application of SSD to the simulation of structural change requires a different semantic association of SD variables with the GIS data model. The commonly used one to one *raster cell-to-stock association* may be replaced by a greater than one to one *raster cells-to-stock ratio*. A stock value represents the area of a structural element such as a specific land cover type. The area exchanged between these stocks is assigned to the raster data model by a set of predefined allocation rules. Processes which modify spatial structures and structures which in turn affect processes define an important feedback component of this system.

As a consequence of this mutual interdependence, interaction between SD and GIS needs to be bidirectional. The tight coupling of the SD software VENSIM to ArcGIS via a Python middleware enables efficient system interactions during program execution. Furthermore, operations performed in GIS and SD need to be synchronized for the structural change model type. The computational overhead associated with these specific coupling features is virtually negligible, as demonstrated in the example of a fictitious system called Daisyworld.

In Daisyworld, a dynamically adapting landscape of black and white daisies balances out temperature changes caused by varying luminosities. The comparison of a spatial Daisyworld simulation to a non-spatial simulation emphasizes the significance of incorporating structural feedbacks inherent in spatial models. The initial composition and land type distribution of the Daisyworld landscape has a consequence on the assignment of future land cover types and simulation results. Moreover, simulation results suggest daisies die to point of extinction in the non-spatial model. This is likely due to the more pronounced response of land cover in the lumped simulation, which doesn't take local growth restrictions (as an expression of spatial structures) into account. Locally restricted daisy growth leads to increased temperatures in the case where luminosity is increasing, whereas the pronounced land cover adaptation in the non-spatial version actually causes decreasing temperatures. Thus, the spatial system exhibits higher robustness, whereas the non-spatial version has higher adaptive capacities.

However, results of the spatial model are not only influenced by the initial landscape structure, but also depend strongly on raster resolution. An increase in spatial resolution goes hand in hand with decreasing adaptive capacities and vice versa. This is due to changes in the raster topology (shared borders between daisy and fertile soil cells).

Nevertheless, results support the idea of utilizing SSD for the simulation of structural change. Compared to diffusion process models, less data links between SD and GIS are required. This is conducive for applications in environments which require large amounts of data. The middleware program can also be used for the implementation of more realistic, multi-disciplinary models in the future.

# 7. Potential Extensions

In the presented approach objects are generated by assigning consistent numeric identifiers to a discrete number of raster cells. Objects assembled in this way are visually interpreted as objects; however they are not fully recognized as objects in GIS. Therefore, the next step towards a more efficient approach is to partially vectorize objects and assign them distinct attributes. Objects metrics may then be analyzed in GIS. This modification of the approach would also allow establishing object hierarchies for multi-scale modeling. Hierarchical organization enables simultaneous simulation processes on multiple scales as well as interactions between them. Moreover, biasing effects of different raster resolutions may be prevented by using a vector model instead.

Possible future applications of the SSD approach include simulating land ownership changes and simulating emergency response during disasters. More specifically, the coupling program is planned to be used for modeling structural changes in grassland agriculture in Austria. This includes simulating competition between different farms to assess ownership changes. In this example structural feedback is considered on a higher resolution (local structures instead of global). Transportation costs are derived for each farm from an infrastructure map. These transportation cost change as the area being farmed changes. In addition, the processes involved operate across multiple scale and domain levels. For instance, climatic variations may be treated as an impact factor on a regional scale, whereas aspects of agrarian economy or policy are typically modeled at the national level.

Moreover, an application of the structural change approach is planned for the dynamic simulation of serviceable emergency response areas in response to natural disasters. Street networks are governed by changing traffic during hazards and access to critical infrastructure may become restricted. Thus, emergency response is functioning as part of a dynamic temporal and spatial landscape. The coupling program aspires to capture these changes and improve disaster planning, response, and recovery efforts.

Additionally, the proposed approach may be especially suitable for applications in the fields of land use and land cover modeling. This is due to the clear distinction between processes and structures as well as the restrictions to qualitative changes in space. The structure change model type may be used as a supplement to commonly used CA and ABM approaches in this field.

# Acknowledgements

The authors appreciate the financial support provided by the Austrian Science Fund (FWF) through the Doctoral College GIScience (DK W 1237-N23) to the first author and the Natural Sciences and Engineering Research Council (NSERC) of Canada CGS doctoral scholarship to the second author.# References